# Higher-order Topological Type-II Hyperbolic Lattices

Liren Chen[1], Jingming Chen[1,*], Zhen Gao[1,†]

[1]State Key Laboratory of Optical Fiber and Cable Manufacturing Technology, Department of Electronic and Electrical Engineering, Guangdong Key Laboratory of Integrated Optoelectronics Intellisense, Southern University of Science and Technology, Shenzhen 518055, China

[†,*]Corresponding author. Email: chenjingming24@163.com; gaoz@sustech.edu.cn (Z.G.)

Recently, higher-order topological phases have been extended from Euclidean lattices to non-Euclidean hyperbolic lattices. Though higher-order topological type-I hyperbolic lattices have been extensively studied, their counterpart, higher-order topological type-II hyperbolic lattices, have never been reported yet. Here, by mapping the celebrated Bernevig–Hughes–Zhang model onto a type-II hyperbolic lattice, we present a theoretical exploration of the first-order topological edge states and second-order topological corner states in a type-II hyperbolic lattice. Compared with the higher-order topological type-I hyperbolic lattices, we discover two unique topological phenomena that stem from the nontrivial geometrical topology of the type-II hyperbolic lattice. First, topological edge and corner states exist on both inner and outer boundaries of the type-II hyperbolic lattice and exhibit higher degeneracy than those in the type-I hyperbolic lattice with only an outer boundary. Second, the degeneracy of type-II hyperbolic corner states can be arbitrarily tuned by changing the characteristic (or inner) radius, in contrast to its type-I counterpart, which is determined by the number of sides of the tessellated polygons. Our work explores topological states in more complex hyperbolic lattices, significantly expanding the research scope of hyperbolic topological physics.

**Introduction**

Topological insulators have become a central paradigm in condensed matter physics [1,2], characterized by their gapped bulk and gapless boundary states, which are protected by a topological invariant [1]. In a $d$-dimensional topological insulator, the bulk–boundary correspondence [3–5] guarantees the existence of topologically protected states on its ($d$-1)-dimensional boundaries, such as edges or surfaces—a hallmark feature of first-order topological phases. Recently, this framework has been extended to higher-order topological phases [6–8], which host topologically protected states on boundaries at least two dimensions lower than the bulk. Specifically, an $n$th-order topological phase in a $d$-dimensional system exhibits topologically protected states localized on ($d$-$n$)-dimensional boundaries. For example, a two-dimensional (2D) second-order topological insulator supports zero-dimensional (0D) corner states, while a three-dimensional (3D) second-order (third-order) topological insulator supports 2D (one-dimensional (1D)) surface (hinge) states.

A defining feature of higher-order topological phases is that their topological protection typically relies on spatial symmetries, primarily rotational symmetry [7,9]. In 2D Euclidean systems, only rotational symmetries compatible with periodic tiling— namely $C_2$, $C_3$, $C_4$, and $C_6$—are possible. Consequently, the degeneracy of second-order topological corner states is restricted to values such as two, three, four, or six, depending on the underlying symmetry [6,9–11]. This restriction raises a natural question: could the richer rotational symmetries available in non-Euclidean geometries give rise to new classes of higher-order topological phases that are impossible in flat-space crystalline systems? This question has motivated growing interest in hyperbolic lattices [12–27]—regular tilings of negatively curved spaces characterized by Schläfli symbols $\{p, q\}$, where $p$ and $q$ represent $q$ copies of $p$-sided polygons meeting at each vertex. Since hyperbolic lattices admit arbitrary $C_p$ rotational symmetries for $p > 2$, they naturally provide an ideal platform to explore higher-order topological phases with corner state degeneracies beyond the limits imposed by Euclidean symmetry classes. Notably, previous works on higher-order type-I hyperbolic lattices have demonstrated topological corner states with degeneracies of 4, 8, 12, or higher, directly determined by the symmetry order $p$ [21,23]. However, the corner state degeneracy of type-I hyperbolic lattices is strictly fixed by the $p$-sided polygons of the tiling without tunability. In contrast, the recently discovered type-II hyperbolic lattices [28,29] introduce a new degree of freedom, the characteristic radius $r_h$, to tune the degeneracy of the hyperbolic topological corner states.

In this work, we map the celebrated Bernevig–Hughes–Zhang (BHZ) model onto a type-II hyperbolic lattice [30] and demonstrate that both its inner and outer boundaries can host first-order topological edge states. By introducing a Wilson mass term to this topological type-II hyperbolic lattice, we obtain a higher-order topological hyperbolic system that supports second-order topological corner states. More interestingly, we find that the degeneracy of these corner states can be controlled by varying the characteristic radius $r_h$, allowing for arbitrary degeneracies that are not limited by the polygonal order $p$. As concrete examples, we demonstrate the emergence of 16-fold and 24-fold degenerate corner states in two distinct higher-order type-II hyperbolic lattices—both tessellated by octagons but characterized by different characteristic radii $r_h$. Our results establish type-II hyperbolic lattices as a versatile platform for realizing tunable, arbitrarily degenerate topological corner states in non-Euclidean geometries.

**Results**
**Quantum spin Hall insulators in type-II hyperbolic lattices**
To construct a topological type-II hyperbolic lattice, we map the BHZ model, also well-known as the quantum spin Hall insulators (QSHIs), onto a type-II hyperbolic lattice, which can be described by the tight-binding model Hamiltonian:

$$H_{\text{QSHI}} = -\frac{1}{2}\sum_{\langle i,j\rangle} c_i^\dagger \left[it_1(\sigma_3 s_1 \cos\theta_{ij} + \sigma_0 s_2 \sin\theta_{ij}) + t_2 \sigma_0 s_3\right] c_j$$
$$+ \sum_i (M + 2t_2) c_i^\dagger \sigma_0 s_3 c_i \qquad (1)$$

where $c_i^\dagger$ and $c_i$ are the creation and annihilation operators of electrons at the site $i$ that have four degrees of freedom, of which the spin and orbital degrees are represented by two sets of Pauli matrices $\sigma_\nu$ and $s_\nu$ with $\nu = 0, 1, 2, 3$, respectively. $M$ denotes the Dirac mass, $t_1$ and $t_2$ are the hopping strengths, $\theta_{ij}$ represents the polar angle of the vector from site $i$ to site $j$, as illustrated in Fig. 1a. Here, we set $t_1 = t_2 = 1$, $M = -1$, and diagonalize the Hamiltonian $H_{\text{QSHI}}$ under open boundary condition. Due to the time-reversal symmetry (TRS) of the Hamiltonian $H_{QSHI}$, as shown in Fig. 1b, the energy spectrum of the Hamiltonian $H_{QSHI}$ of topological type-II hyperbolic lattice exhibit double degenerate inner and outer helical edge states (red dots) in the bulk energy gap (orange region), in contrast to the topological type-I hyperbolic lattices which only support topological edge states on its outer boundary [16–19,25,26]. Note that the energy values of the inner and outer helical edge states are slightly different (see inset in Fig. 1b), arising from the difference in $\theta_{ij}$ between sublattices inside and outside $r_h$ in the Hamiltonian $H_{\text{QSHI}}$.

To characterize the topological nature of the type-II hyperbolic edge states, we compute the spin Bott index $C_B$ (see details in Appendix A) as a function of energy [31,32]. As illustrated in Fig. 1c, the index has a value of $C_B = -1$ precisely in the energy gap where the edge states reside, whereas it vanishes ($C_B = 0$) within the bulk bands. This behavior confirms the nontrivial band topology of the topological hyperbolic edge states. Fig. 1d displays the intensity distributions of the topological hyperbolic edge states on the outer (upper panel) and inner (lower panel) boundaries. Here, we adopt a point source (green star) to excite the topological hyperbolic edge states, and an absorbing boundary condition (black rectangle) is applied at each respective edge. The spin-up and spin-down components of the helical edge states are represented by green and orange color scales, respectively, highlighting their spatial separation and chiral propagation along the curved edges.

**Higher-order topological insulators in type-II hyperbolic lattices**
Besides the first-order topological edge states, topological type-II hyperbolic lattices can also support second-order topological corner states with unique properties compared with their type-I counterparts [19,21,23]. To construct the type-II hyperbolic higher-order topological insulators (HOTIs), we introduce a Wilson mass

term into the aforementioned type-II hyperbolic QSHIs, and the system's Hamiltonian becomes:

$$H_{\text{HOTI}} = H_{\text{QSHI}} + H_{\text{m}}, \qquad H_{\text{m}} = \frac{g}{2}\sum_{\langle i,j\rangle} c_i^\dagger \cos(\frac{k}{2}\theta_{ij})\,\sigma_1 s_1 c_j \qquad (2)$$

where $g$ represents the magnitude of the Wilson mass. The additional mass term $H_{\text{m}}$ breaks the TRS in the Hamiltonian $H_{\text{QSHI}}$ and gaps the one-dimensional (1D) helical edge states in the first-order QSHIs. Since this additional mass term periodically changes its sign along the outer and inner boundaries, $k$ (rotation symmetry of the type-II hyperbolic lattice) mass domain walls are formed at the locations where the mass flips its sign, inducing a type-II hyperbolic HOTI with zero-dimensional (0D) topological corner states at the inner and outer boundaries.

To demonstrate the existence of type-II hyperbolic zero-energy corner states (ZECSs), we calculate the eigenenergy and eigenstates of the Hamiltonian $H_{HO}$ in real space with open boundary conditions. The energy spectrum of the Hamiltonian $H_{HO}$ ($k=4$) with different Wilson mass magnitudes $g$ (Fig. 2a) shows the existence of eight-fold degenerate ZECSs (see inset in Fig. 2a) in the bulk bandgap when $0 < g < 1.45$ (horizontal red solid line), including four (four) ZECSs on the outer (inner) boundary. To demonstrate the localization of the ZECSs, we compute the zero-energy LDOS defined as $\rho(0,\boldsymbol{r}) = \sum_{i,j}\delta(0-E_i)\,|\Psi_{i,j}(\boldsymbol{r})|^2$, where $\Psi_{i,j}(\boldsymbol{r})$ is the $j$th component of the $i$th eigenstate at site $\boldsymbol{r}$. Then we take an average of the zero-energy LDOS over the site collection $\Pi_S$ as $\bar{\rho} = (1/N)\sum_{\boldsymbol{r}\in\Pi_S}\rho(0,\boldsymbol{r})$ near the corner (outer/inner boundary site with polar angle $\theta = \pi/4$), edge (collection of outer/inner boundary sites between two nearest neighboring corners) and bulk (collection of bulk sites on the characteristic radius) with different $g$, as shown in Fig. 2b. When $g = 0$, the type-II hyperbolic lattice is in the first-order topological phase with gapless helical edge states. As $g$ increases from zero, the energy gaps at the outer and inner boundaries are opened, resulting in the outer and inner corner states, reflected by the increase (decrease) of the corner (edge) averaged LDOS and almost vanished bulk averaged LDOS. When $g$ approaches $g_T = 1.45$, the bulk band gap gradually closes and the corner (bulk) averaged LDOS decrease (increase). To further identify the topological phase transition in this process, we calculate the real-space quadrupole moment $Q_{xy}$ as a function of $g$ (see details in Appendix B), as shown in Fig. 2c. We find that $Q_{xy}$ changes from 0.5 to 0 at $g = g_T = 1.45$, indicating a topological phase transition from a higher-order topological phase to a trivial phase, perfectly aligning with the interval in which the ZECSs exist in Fig. 2a.

To directly visualize the localized corner states, we again place four localized excitation point source (green star) at the outer and inner boundary, Fig. 2d shows the density distribution of the corner states at the two boundrays when $g = g_1 = 0.5$, where the dominated states are inner and outer corner states indicated in Fig. 2b. As presented in the graph, we can see that both the outer and inner corner states are well localized at the four corners near the excitation sources, and there is no leakage to the edge and bulk.

**Arbitrarily degenerate corner states via tuning the characteristic radius of the type-II hyperbolic lattice**

More interestingly, the rotational symmetry of type-II hyperbolic lattice, which is parameterized by $k$, can be arbitrarily tuned by changing its characteristic radius $r_h$ (or inner radius $r_{in}$). This relation is governed by the equation $r_h = e^{-2\pi/kP}$. Unlike the conventional type-I hyperbolic lattices, where the degeneracy of ZECSs is constrained by the $p$-sided polygons used for tessellation. The characteristic radius offers a new degree of freedom to tune corner mode degeneracies that are not restricted by the polygon's geometry.

To investigate the effect of $r_h$ on the degeneracy of ZECSs in type-II hyperbolic lattice, we firstly calculate the degeneracy of ZECSs as a function of the characteristic radius $r_h$, as shown in Fig. 3a. The results demonstrate that the degeneracy of corner states in type-II hyperbolic lattices can be arbitrarily increased by increasing $r_h$, unlike the type-I hyperbolic lattices, where the degeneracy of corner states is solely determined by the number of sides of the tessellated polygons [19,21,23]. Furthermore, the inset in Fig. 3a shows that the degeneracy of corner states increases in direct proportion to the rotational symmetry parameter $k$ and is exactly twice the number of $k$. It reflects the direct relation between the degeneracy of the ZECSs and the rotation symmetry of the type-II hyperbolic lattice which can be tuned by changing its characteristic radius, confirming the tunable nature of the type-II hyperbolic corner states.

To illustrate this tunable degeneracy, we take two octagon tessellated type-II hyperbolic lattices as examples, corresponding to points 1 and 2 in Fig. 3a. At $r_h = 0.604$ (point 1), the energy spectrum in Fig. 3b reveals 16 ZECSs, consistent with the calculated degeneracy. The intensity distribution of these ZECSs, as shown in Fig. 3d, clearly displays 8 degenerate corner modes at both the inner and outer boundaries of the lattice. This confirms the expected 16-fold degeneracy. We then increase the characteristic radius to $0.715$ (point 2), as shown in Fig. 3a. The energy spectrum and intensity distribution in Figs. 3c and 3e show an increased degeneracy to 24, with 12 degenerate corner modes appearing at both inner and outer boundaries. All these degenerate corner modes arise from the rotation symmetry of the type-II hyperbolic lattices. But different from the previously studied type-I hyperbolic lattices, now the rotation symmetry can be controlled by changing the characteris radius $r_h$, offering a novel way to tune the degeneracy of higher-order topological corner states.

**Conclusion**

In conclusion, we have theoretically explored the nontrivial first-order topological edge states and higher-order topological corner states in a newly discovered type-II hyperbolic lattice. By mapping the BHZ model to the type-II hyperbolic lattice, we demonstrate that both the inner and outer boundaries of type-II hyperbolic lattices can support the first-order helical edge states and second-order topological corner states, exhibiting a higher degree of degeneracy than their type-I counterparts which only exist on the outer boundary. furthermore, we demonstrate that the degeneracy (number) of the type-II hyperbolic corner states can be arbitrarily tuned by changing the characteristic (or inner) radius, in contrast to their type-I counterparts whose number is determined by the number of sides of the tessellated polygons.

Given current experimental platforms, the topological type-II hyperbolic lattices can be realized in circuit quantum electrodynamics [33], electric circuits [19,20,24,34], nonreciprocal networks [25] and integrated nanophotonic chips [26]. Since the type-II hyperbolic lattices possess both inner and outer boundaries supporting counterpropagating topological edge states, we envision that it can provide an ideal platform to extend many important topological phenomena such as Thouless pumping [35,36], Laudal-Zener transition [37], and Laughlin's pump [38–40] from Euclidean to non-Euclidean geometries. Our work thus establishes type-II hyperbolic lattices as a versatile and experimentally accessible platform for exploring novel topological physics beyond flat space.

**Appendix A: Calculation of spin Bott index**

In periodic QSHI, the topogical nature of the edge states is captured by a $\mathbb{Z}_2$ invariant. However, in a type-II hyperbolic lattice that periodicity is absent, we instead need to employ the spin Bott index as a marker for topological transition. Below we follow the construction of Ref. [32] and briefly summarize the calculation steps.

We first introduce two individual spin vectors $C_{B_+}$ and $C_{B_-}$ given by

$$C_{B_\pm} = \frac{1}{2\pi} \text{Im}\{tr[\log(V_\pm U_\pm V_\pm^\dagger U_\pm^\dagger)]\} \quad (A1)$$

Where "+" and "−" label the spin-up and spin-down parts. Respectively, the spin Bott index is then

$$C_B = \frac{1}{2}(C_{B_+} - C_{B_-}) \quad (A2)$$

The projected position operator $U_\pm$ and $V_\pm$ are built from the projector $P_\pm$ onto the occupied spin-up or spin-down subspace: $U_\pm = P_{\pm\Sigma} e^{i2\pi X} P_\pm + (I - P_\pm)\Sigma\_\_$ and $V_\pm = P_{\pm\Sigma} e^{i2\pi Y} P_\pm + (I - P_\pm)$. Here $X$ and $Y$ are diagonal matrices whose entries $x_i, y_i$ are the rescaled real-space coordinates of the $i$th lattice site. Concretely, an original coordinate $x_i, y_i \in$ (-1, 1) is mapped to the unit interval $x_i, y_i \in$ (0, 1). $P_\pm = \sum_n^{N_{occ}/2} |\pm\psi_n\rangle\langle\pm\psi_n|$ is the projector operator that projectes onto the occupied spin up or down eigenstates $|\pm\psi_n\rangle$. It has already been shown in Ref. [32] that, for any QSHI with or without periodicity, the spin Bott index $C_B$ is equivalent to the standard $\mathbb{Z}_2$ invariant.

**Appendix B: Calculation of real-space quadrupole moment**

In conventional flat lattices, higher-order topological states are usually characterized by quadrupole moments in momentum space. Since there is no well-defined translation symmetry in the type-II hyperbolic lattice, we utilize the real-space quadrupole moment [23,41,42] to characterize the topological properties of the type-II hyperbolic zero-energy corner states (ZECSs). The real-space quadrupole moment $Q_{xy}$ is defined as:

$$Q_{xy} = \left[\frac{1}{2\pi}\text{Im}\log\det(\Psi_{\text{occ}}^\dagger U \Psi_{\text{occ}}) - Q_0\right] \mod 1, \tag{B1}$$

with $Q_0 = \frac{1}{2}\sum_j x_j y_j / A$, where $\Psi_{\text{occ}}$ is the occupied eigenstates of $H_{\text{HOTI}}$, $U$ is a diagonal matrix with elements $e^{2\pi i x_j y_j / A}$, $(x_j, y_j)$ is the new coordinate of the jth site after translating the coordinate interval $x_j, y_j \in$ (-1, 1) to $x_j, y_j \in$ (0, 1), A is the Euclidean area of Poincaré ring.

**Data availability**
All data that support the plots within this paper and other findings of this work are available from the corresponding authors upon reasonable request.

**Code availability**
All the codes used to generate and/or analyze the data in this work are available from the corresponding authors upon reasonable request.

**Reference**

[1]  A. Bansil, H. Lin, and T. Das, Colloquium: Topological band theory, Rev. Mod. Phys. **88**, 021004 (2016).

[2]  J. E. Moore, The birth of topological insulators, Nature **464**, 7286 (2010).

[3]  C. L. Kane and E. J. Mele, $Z_2$ Topological Order and the Quantum Spin Hall Effect, Phys. Rev. Lett. **95**, 146802 (2005).

[4]  L. Fu, C. L. Kane, and E. J. Mele, Topological Insulators in Three Dimensions, Phys. Rev. Lett. **98**, 106803 (2007).

[5]  L. Fu and C. L. Kane, Topological insulators with inversion symmetry, Phys. Rev. B **76**, 045302 (2007).

[6]  W. A. Benalcazar, B. A. Bernevig, and T. L. Hughes, Quantized electric multipole insulators, Science **357**, 61 (2017).

[7]  F. Schindler, A. M. Cook, M. G. Vergniory, Z. Wang, S. S. P. Parkin, B. A. Bernevig, and T. Neupert, Higher-order topological insulators, Sci. Adv. **4**, eaat0346 (2018).

[8]  B. Xie, H.-X. Wang, X. Zhang, P. Zhan, J.-H. Jiang, M. Lu, and Y. Chen, Higher-order band topology, Nat Rev Phys **3**, 520 (2021).

[9]  X. Ni, M. Weiner, A. Alù, and A. B. Khanikaev, Observation of higher-order topological acoustic states protected by generalized chiral symmetry, Nature Mater **18**, 113 (2019).

[10] X.-L. Sheng, C. Chen, H. Liu, Z. Chen, Z.-M. Yu, Y. X. Zhao, and S. A. Yang, Two-Dimensional Second-Order Topological Insulator in Graphdiyne, Phys. Rev. Lett. **123**, 256402 (2019).

[11] C. Chen, Z. Song, J.-Z. Zhao, Z. Chen, Z.-M. Yu, X.-L. Sheng, and S. A. Yang, Universal Approach to Magnetic Second-Order Topological Insulator, Phys. Rev. Lett. **125**, 056402 (2020).

[12] J. Maciejko and S. Rayan, Hyperbolic band theory, Sci. Adv. **7**, eabe9170 (2021).

[13] J. Maciejko and S. Rayan, Automorphic Bloch theorems for hyperbolic lattices, Proc. Natl. Acad. Sci. USA **119**, e2116869119 (2022).

[14] N. Cheng, F. Serafin, J. McInerney, Z. Rocklin, K. Sun, and X. Mao, Band Theory and Boundary Modes of High-Dimensional Representations of Infinite Hyperbolic Lattices, Phys. Rev. Lett. **129**, 088002 (2022).

[15] P. M. Lenggenhager, J. Maciejko, and T. Bzdušek, Non-Abelian Hyperbolic Band Theory from Supercells, Phys. Rev. Lett. **131**, 226401 (2023).

[16] S. Yu, X. Piao, and N. Park, Topological Hyperbolic Lattices, Phys. Rev. Lett. **125**, 053901 (2020).

[17] Z.-R. Liu, C.-B. Hua, T. Peng, and B. Zhou, Chern insulator in a hyperbolic lattice, Phys. Rev. B **105**, 245301 (2022).

[18] D. M. Urwyler, P. M. Lenggenhager, I. Boettcher, R. Thomale, T. Neupert, and T. Bzdušek, Hyperbolic Topological Band Insulators, Phys. Rev. Lett. **129**, 246402 (2022).

[19] W. Zhang, H. Yuan, N. Sun, H. Sun, and X. Zhang, Observation of novel topological states in hyperbolic lattices, Nat Commun **13**, 2937 (2022).

[20] A. Chen, H. Brand, T. Helbig, T. Hofmann, S. Imhof, A. Fritzsche, T. Kießling, A.

Stegmaier, L. K. Upreti, T. Neupert, et al., Hyperbolic matter in electrical circuits with tunable complex phases, Nat Commun **14**, 622 (2023).

[21] Z.-R. Liu, C.-B. Hua, T. Peng, R. Chen, and B. Zhou, Higher-order topological insulators in hyperbolic lattices, Phys. Rev. B **107**, 125302 (2023).

[22] J. Sun, C.-A. Li, S. Feng, and H. Guo, Hybrid higher-order skin-topological effect in hyperbolic lattices, Phys. Rev. B **108**, 075122 (2023).

[23] Y.-L. Tao and Y. Xu, Higher-order topological hyperbolic lattices, Phys. Rev. B **107**, 184201 (2023).

[24] W. Zhang, F. Di, X. Zheng, H. Sun, and X. Zhang, Hyperbolic band topology with non-trivial second Chern numbers, Nat Commun **14**, 1083 (2023).

[25] Q. Chen, Z. Zhang, H. Qin, A. Bossart, Y. Yang, H. Chen, and R. Fleury, Anomalous and Chern topological waves in hyperbolic networks, Nat Commun **15**, 2293 (2024).

[26] L. Huang, L. He, W. Zhang, H. Zhang, D. Liu, X. Feng, F. Liu, K. Cui, Y. Huang, W. Zhang, et al., Hyperbolic photonic topological insulators, Nat Commun **15**, 1647 (2024).

[27] T. Tummuru, A. Chen, P. M. Lenggenhager, T. Neupert, J. Maciejko, and T. Bzdušek, Hyperbolic Non-Abelian Semimetal, Phys. Rev. Lett. **132**, 206601 (2024).

[28] J. Chen, F. Chen, L. Yang, Y. Yang, Z. Chen, Y. Wu, Y. Meng, B. Yan, X. Xi, Z. Zhu, et al., AdS/CFT Correspondence in Hyperbolic Lattices, arXiv:2305.04862.

[29] J. Chen, L. Yang, and Z. Gao, Dynamic transfer of chiral edge states in topological type-II hyperbolic lattices, Commun Phys **8**, 97 (2025).

[30] B. A. Bernevig, T. L. Hughes, and S.-C. Zhang, Quantum Spin Hall Effect and Topological Phase Transition in HgTe Quantum Wells, Science **314**, 1757 (2006).

[31] H. Huang and F. Liu, Quantum Spin Hall Effect and Spin Bott Index in a Quasicrystal Lattice, Phys. Rev. Lett. **121**, 126401 (2018).

[32] H. Huang and F. Liu, Theory of spin Bott index for quantum spin Hall states in nonperiodic systems, Phys. Rev. B **98**, 125130 (2018).

[33] A. J. Kollár, M. Fitzpatrick, and A. A. Houck, Hyperbolic lattices in circuit quantum electrodynamics, Nature **571**, 45 (2019).

[34] P. M. Lenggenhager, A. Stegmaier, L. K. Upreti, T. Hofmann, T. Helbig, A. Vollhardt, M. Greiter, C. H. Lee, S. Imhof, H. Brand, et al., Simulating hyperbolic space on a circuit board, Nat Commun **13**, 4373 (2022).

[35] D. J. Thouless, Quantization of particle transport, Phys. Rev. B **27**, 6083 (1983).

[36] R. Citro and M. Aidelsburger, Thouless pumping and topology, Nat Rev Phys **5**, 87 (2023).

[37] Z.-G. Chen, W. Tang, R.-Y. Zhang, Z. Chen, and G. Ma, Landau-Zener Transition in the Dynamic Transfer of Acoustic Topological States, Phys. Rev. Lett. **126**, 054301 (2021).

[38] R. B. Laughlin, Quantized Hall conductivity in two dimensions, Phys. Rev. B **23**, 5632 (1981).

[39] A. Fabre, J.-B. Bouhiron, T. Satoor, R. Lopes, and S. Nascimbene, Laughlin's Topological Charge Pump in an Atomic Hall Cylinder, Phys. Rev. Lett. **128**,

173202 (2022).

[40] M. Kawamura, M. Mogi, R. Yoshimi, T. Morimoto, K. S. Takahashi, A. Tsukazaki, N. Nagaosa, M. Kawasaki, and Y. Tokura, Laughlin charge pumping in a quantum anomalous Hall insulator, Nat. Phys. **19**, 333 (2023).

[41] W. A. Wheeler, L. K. Wagner, and T. L. Hughes, Many-body electric multipole operators in extended systems, Phys. Rev. B **100**, 245135 (2019).

[42] C.-A. Li, B. Fu, Z.-A. Hu, J. Li, and S.-Q. Shen, Topological Phase Transitions in Disordered Electric Quadrupole Insulators, Phys. Rev. Lett. **125**, 166801 (2020).

**Acknowledgments**

Z.G. acknowledges funding from the National Key R&D Program of China (grant no. 2025YFA1412300), National Natural Science Foundation of China (grants no. 62361166627 and 62375118), Guangdong Basic and Applied Basic Research Foundation (grant no. 2024A1515012770), Shenzhen Science and Technology Innovation Commission (grants no. 202308073000209), and High-level Special Funds (grant no. G03034K004).

**Competing Interests**

The authors declare no competing interests.

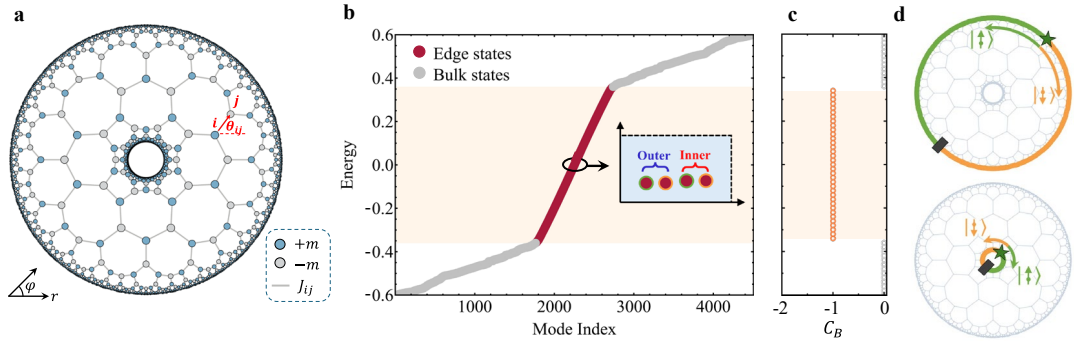

**Fig. 1 | Type-II hyperbolic QSHIs. (a)** Schematic illustration of the BHZ model on a type-II hyperbolic {0.365, 8, 3} lattice. **(b)** Energy spectrum of Hamiltonian $H_{\text{QSHI}}$ in the type-II hyperbolic. Inset displays the zoom-in view of the in-gap inner and outer edge states near zero energy. **(c)** Spin Bott index $C_B$ as a function of energy. **(d)** Intensity distributions of the outer (upper panel) and inner (lower panel) helical edge states excited by edge sources (green stars), corresponding to the four edge states marked in **(a)**. An absorber (black block) is adopted to identify the helicity of the edge states.

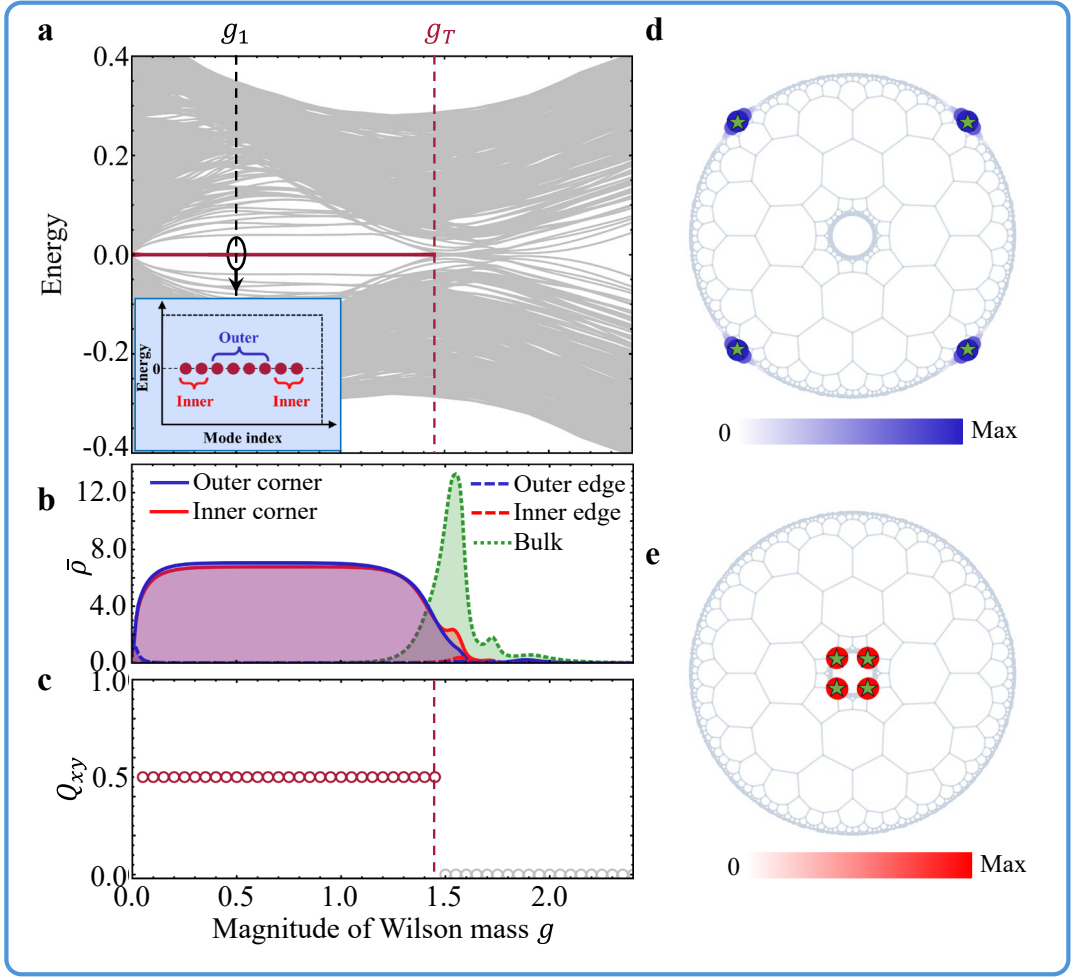

**Fig. 2 | Type-II hyperbolic HOTIs**. (a) Energy spectrum of Hamiltonian $H_{HOTI}$ in the type-II hyperbolic {0.365, 8, 3} lattice with different magnitudes of Wilson mass $g$. The horizontal red line represents the ZECSs. The black and red vertical dashed lines indicate three characteristic values of $g$. Inset displays the zoom-in view of the eight-fold degenerate ZECSs. (b) Zero-energy averaged LDOS $\bar{\rho}$ at the positions near an outer corner (blue line), an inner corner (red line), a part of outer edge between two neighboring outer corners (light blue line), a part of inner edge between two neighboring inner corners (light red line), and the bulk sites on the characteristic radius (green line) as a function of $g$. (c) Quadrupole moment $Q_{xy}$ as a function of $g$. (d) and (e) Intensity distributions of outer (upper panel) and inner (lower panel) ZECSs excited by the sources (green stars) with $g = g_1 = 0.5$.

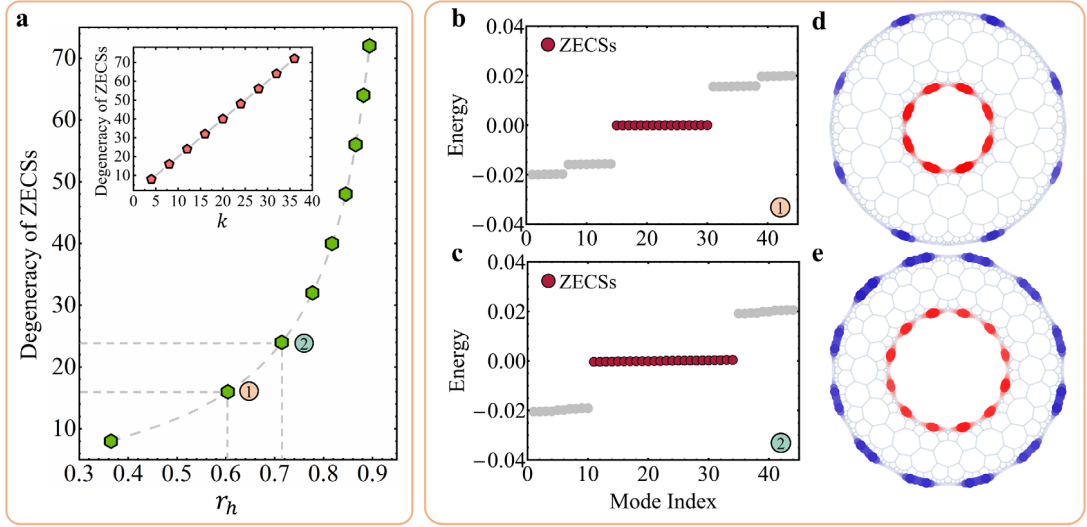

**Fig. 3 | Arbitrarily tunable degeneracy of ZECSs by changing the characteristic radius $r_h$.** (a) The degeneracy of ZECSs as a function of the characteristic radius $r_h$. Inset displays the degeneracy of ZECSs with different rotation symmetries $k$. (b) and (c) Energy spectra for $H_{HOTI}$ on two type-II hyperbolic lattices with (b) $r_h = 0.604$, $k = 8$ and $g = 0.5$ and (c) $r_h = 0.715$, $k = 12$ and $g = 1.1$, respectively. (d) and (e) Intensity distributions of (d) 16-fold and (e) 24-fold degenerate type-II hyperbolic corner states, corresponding to the ZECSs (red dots) in b and c, respectively.